\documentclass[a4paper]{JAC2003}
\title{COHERENT BEAM--BEAM EXPERIMENTS AND IMPLICATIONS FOR HEAD-ON COMPENSATION\thanks{Work supported by Brookhaven Science Associates, LLC under Contract
No. DE-AC02-98CH10886 with the U.S. Department of Energy, and in part by the U.S. LHC Accelerator Research Program.}}

\author{S. White, M. Blaskiewicz, W. Fischer, Y. Luo \\
{BNL, Upton, NY, USA}}

\usepackage{graphicx}
\usepackage{booktabs}
\usepackage{varwidth}
\usepackage{xcolor}

\begin{document}
\maketitle

\begin{abstract}
In polarized proton operation in the Relativistic Heavy Ion Collider (RHIC) coherent beam--beam modes are routinely observed with beam transfer function
measurements. These modes can become unstable under external excitation or in the presence of impedance.
This becomes even more relevant in the presence of head-on compensation, which reduces the beam--beam tune spread
and hence Landau damping. We report on experiments and simulations carried out to understand the impact of
coherent modes on operation with electron lenses.
\end{abstract}

\section{Introduction}

The Relativistic Heavy Ion Collider (RHIC) is currently operating between the 2/3 and 7/10 resonances with a beam--beam parameter of approximately
0.015 leaving little space for significant increase in luminosity. The RHIC luminosity upgrade program \cite{RHIC_upgrade}
aims at an increase of the luminosity by a factor of 2. In order to accommodate the significant increase in beam--beam
tune spread it was decided to install electron lenses to compensate for the beam--beam non-linearities and effectively reduce
the tune spread at constant bunch intensity. This technology was first developed at the Tevatron where it was tested for head-on
compensation \cite{tev} and then successfully used for long-range compensation, abort gap cleaning \cite{tev_ag} and collimation
studies \cite{tev_hollow}.

\begin{figure}[htb]
\begin{center}
\includegraphics[width=0.45\textwidth]{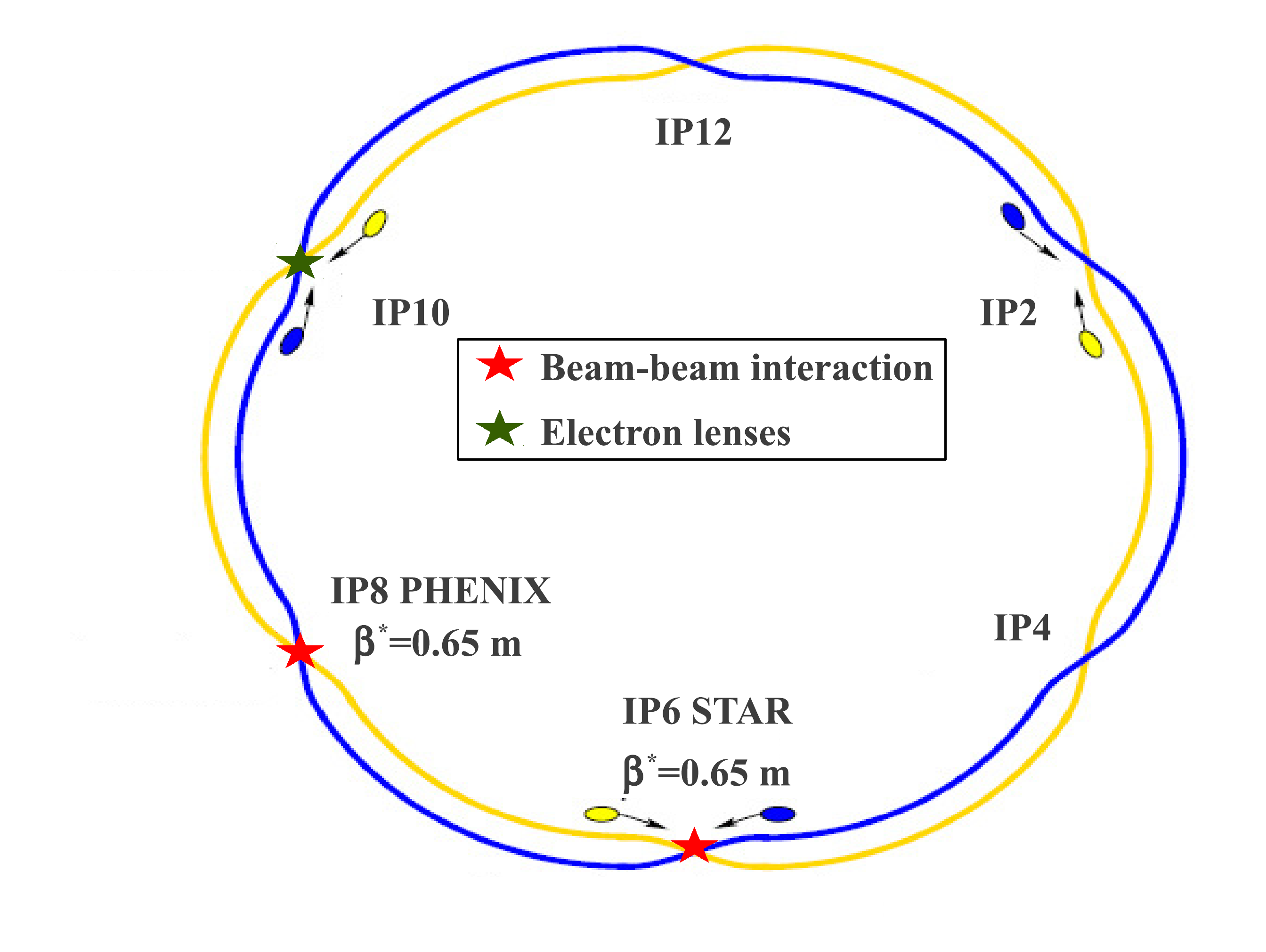}
\end{center}
\caption{Layout of the RHIC collider. The colliding IPs are denoted by the red stars, the head-on compensation by the green star.}
\label{RHIC}
\end{figure}

The RHIC collider consists of two rings where the beams are colliding in interaction points IP6 and IP8 as shown in Fig. \ref{RHIC}. The two electron lenses,
one for each ring, will be located close to IP10. Studies regarding dynamic aperture were performed and showed improvements for high
beam--beam parameter \cite{RHIC_DA}. The details about the status and construction of the electron lens can be found in Ref. \cite{status}.
These simulations however did not cover the coherent beam--beam effects related to the electron lens. The failure in increasing the
luminosity in the DCI (Dispositif de Collisions dans l'Igloo) four-beam experiment ($e^+e^-e^+e^-$) was attributed to coherent effects \cite{DCI}, which should therefore be carefully
investigated. This paper reports on strong--strong beam--beam simulations performed using the RHIC lattice and upgrade parameters and related beam experiments
to understand the impact of the coherent beam--beam effects in the presence of electron lenses.

\section{Model}

The simulation code BeamBeam3D \cite{BB3D} was used for this study. BeamBeam3D is a fully parallelized three-dimensional code allowing for self-consistent
field calculation of arbitrary distributions and tracking of multiple bunches. The transport from one IP to the other is done through linear transfer maps.
The beam fields are calculated by solving the Poisson equation using a shifted integrated Green function method which is efficiently computed with a
FFT-based algorithm on a uniform grid.

In order to correctly model the RHIC lattice the Twiss parameters are extracted at each IP, including the one where the head-on compensation takes place,
and used to compute the transfer maps. As shown in Fig. \ref{RHIC}, the symmetry of the different colliding IPs allows one to reduce the number of bunches to
three per beam to simulate the full collision pattern.

The electron lens is modelled as a thin-lens Gaussian beam located exactly at IP10 for both beams. The size of the electron beam is determined by the
lattice parameters. The phase advance between IP8 and IP10 is set exactly to $\pi$ by artificially shifting the phase between these two IPs and evenly
compensating the global tune change with the other arcs.

\begin{figure}[htb]
\begin{center}
\includegraphics[width=0.45\textwidth]{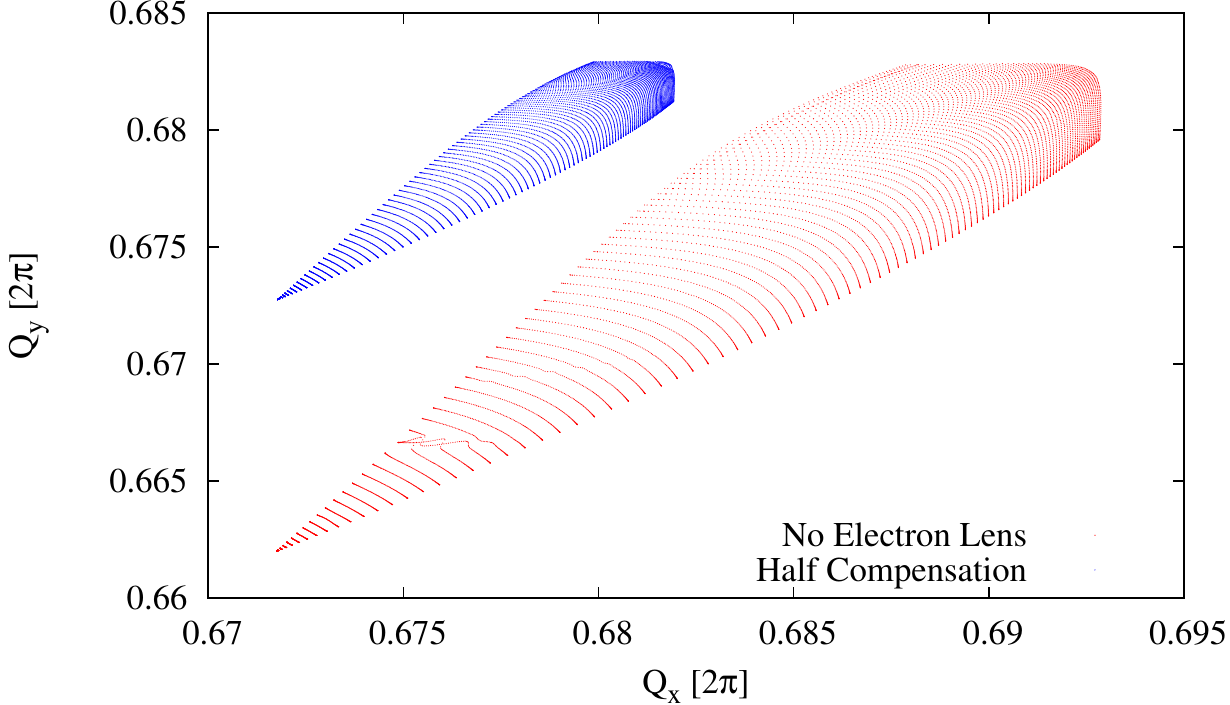}
\end{center}
\caption{Tune footprint computed with BeamBeam3D for an intensity of $3.0\times10^{11}$ protons per bunch with and without compensation. No compensation in red, half compensation in blue.}
\label{fp}
\end{figure}

Figure \ref{fp} shows the footprints calculated with BeamBeam3D for an intensity of $3.0\times10^{11}$ protons per bunch without compensation and with half
compensation. The footprint with compensation was artificially shifted for better visibility. As expected, we observe a reduction of the tune spread by a factor of 2.
One can also see that the footprint without compensation is crossing the $3Q_y$ resonance, indicating that the machine cannot be operated with such high beam--beam
parameter.

\section{Coherent beam--beam simulations}

In addition to the single-particle effects described in the previous sections, colliding beams will experience coherent dipole
oscillation driven by the beam--beam force.
In the simplest case of one interaction point two main modes arise corresponding to the two bunches
oscillating in phase ($\sigma$-mode) or out of phase ($\pi$-mode). The $\sigma$-mode will oscillate at the betatron
frequency and the $\pi$-mode will be shifted, negatively for equally charged beams, with respect to the $\sigma$-mode
by an amount $Y\cdot\xi$, where $Y$ is the Yokoya factor and $\xi$ the beam--beam parameter \cite{Yokoya}. 

The collision pattern at RHIC can be reduced to three colliding bunches theoretically giving rise to six coherent dipole modes.
In reality, only two modes are observed as the other ones are located inside or very close to the incoherent tune spread and
are Landau damped.

\begin{figure}[htb]
\begin{center}
\includegraphics[width=0.45\textwidth]{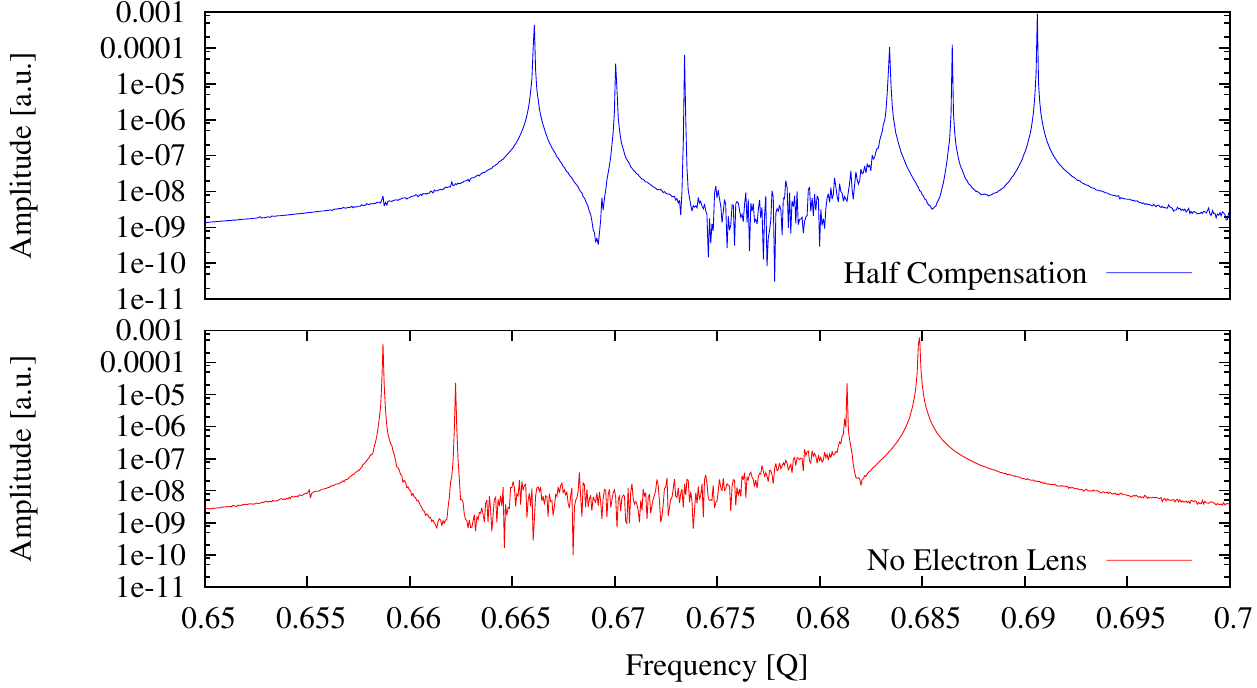}
\end{center}
\caption{Simulated coherent modes with (top) and without (bottom) half compensation with a bare lattice tune of 0.685.}
\label{modes}
\end{figure}

Figure \ref{modes} shows a strong--strong simulation of the RHIC lattice with and without compensation. The bare lattice
tunes used for this simulation are (0.695, 0.685) as defined in the design and the beam--beam parameter $\xi$ per IP is
0.011. Only the vertical plane is shown but a similar picture is observed in the horizontal plane. The coherent modes are excited
with an initial kick of 0.1~$\sigma$. 

As predicted by the weak--strong simulations in Fig. \ref{fp}, the incoherent continuum is reduced by the head-on compensation.
The bare lattice tunes, or $\sigma$-mode in this plot, are shifted by $\xi/2\approx0.005$ corresponding to the coherent beam--beam tune shift induced by the
quadrupolar part of the beam--beam force. This effect can be easily predicted and corrected for.
The phase advance between IPs is also modified leading to slightly different relative frequencies of the modes.

In the presence of head-on compensation, the distance in tune space covered by the coherent modes therefore remains approximately constant while the incoherent tune
spread is significantly reduced. All six coherent modes are now observed as they are moved out of the continuum and not Landau
damped any more. Head-on compensation with electron lenses reduces the intrinsic stabilizing properties of the beam--beam interaction. This could
give rise to coherent dipole instabilities driven by external sources of excitation or impedance.

\section{Effect of the 2/3 resonance on coherent modes}

As seen in Fig. \ref{modes}, even if the incoherent tune spread is reduced, the tune space covered by the coherent modes remains constant and
will overlap the 2/3 resonance in the case of the RHIC working point. While it is difficult to experimentally reproduce the reduction of the tune
spread induced by the electron lenses, we verified experimentally that driving the $\pi$-mode onto this resonance would not excite coherent
dipole motion or degrade the beam lifetime.

\begin{figure}[htb]
\begin{center}
\includegraphics[width=0.45\textwidth]{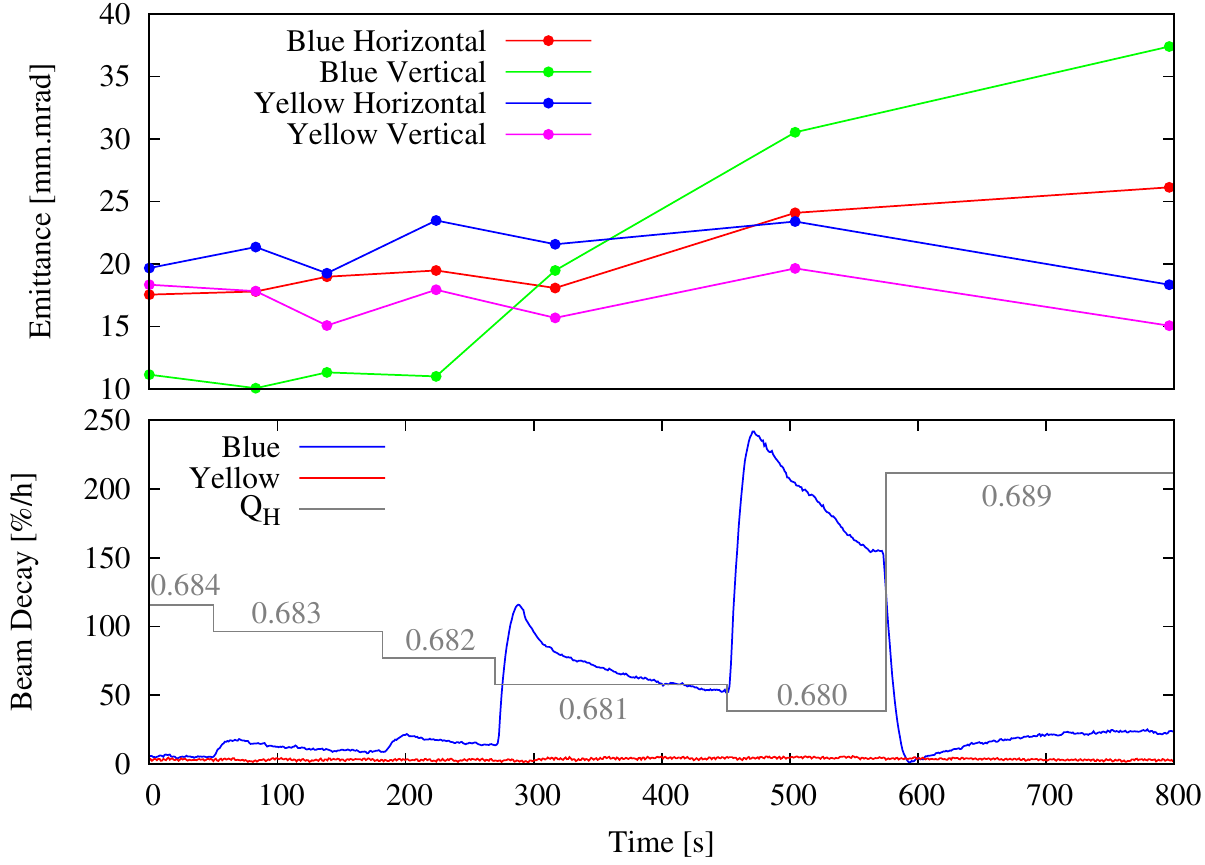}
\end{center}
\caption{Tune scan towards the 2/3 resonance with colliding beams. The top plot shows the emittance during the scan and
the bottom plot the beam decay. The tunes were reconstructed using measurements with non-colliding beams.}
\label{qscan}
\end{figure}

\begin{figure}[htb]
\begin{center}
\includegraphics[width=0.45\textwidth]{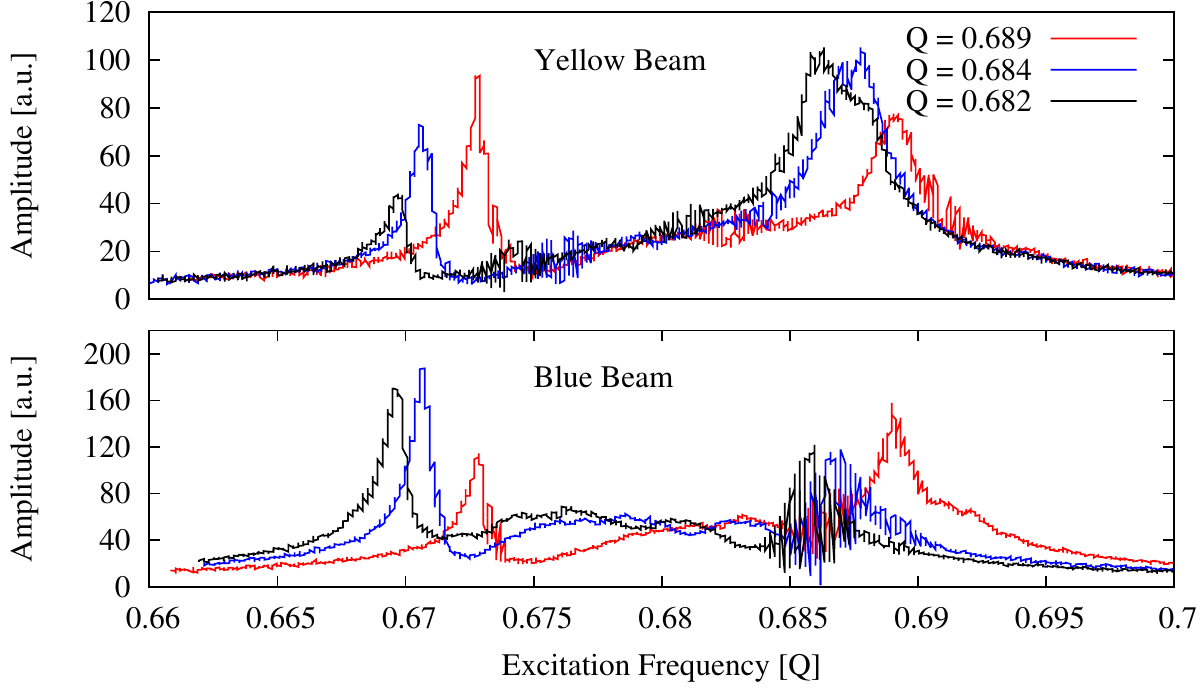}
\end{center}
\caption{BTF (Beam Transfer Function) data during the tune scan towards the 2/3 resonance.}
\label{btf_qscan}
\end{figure}

For this experiment we moved only the two tunes of the Blue beam towards the 2/3 resonance keeping the difference $Q_x-Q_y=0.004$, see Figs. \ref{qscan} and \ref{btf_qscan}.
This was done with a beam--beam parameter estimated to be 0.011. The onset of losses is observed at (0.687, 0.683); at these tunes the location of the $\pi$-mode
is 0.669 and the zero-amplitude particles are at 0.672: no emittance blow-up is observed at that point. Losses are observed only in the Blue beam, indicating
that the $\pi$-mode, which has the same frequency for both beams, is insensitive to the 2/3 resonance. The stop band of the 2/3 resonance with non-colliding
beams was estimated to be around 0.005, which is consistent with losses of low-amplitude particles in our case. As we moved the beam closer to the resonance
strong losses associated with emittance blow-up were observed only in the Blue beam. In addition, no unusual activity was observed in the tune spectrum during
the whole experiment, pointing towards a reduction of the dynamic aperture rather than the excitation of coherent modes.

\begin{figure}[htb]
\begin{center}
\includegraphics[width=0.45\textwidth]{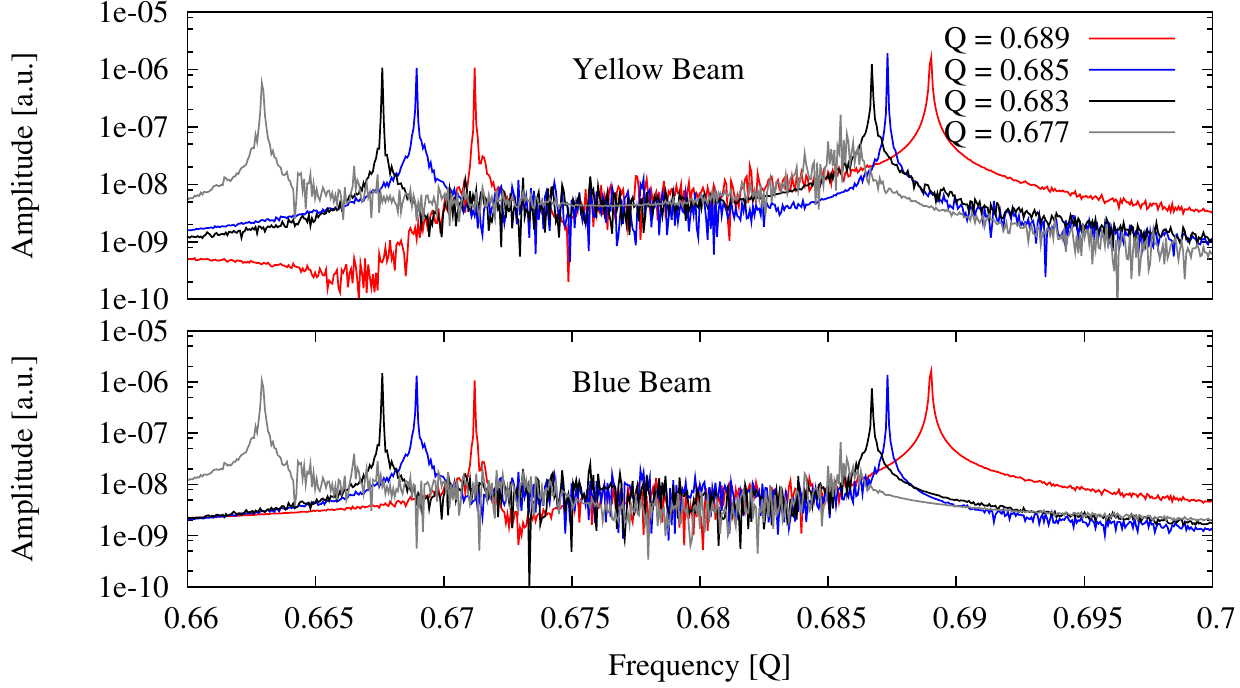}
\end{center}
\caption{Simulated spectrum reproducing the experiment.}
\label{fft_qscan}
\end{figure}

\begin{figure}[htb]
\begin{center}
\includegraphics[width=0.45\textwidth]{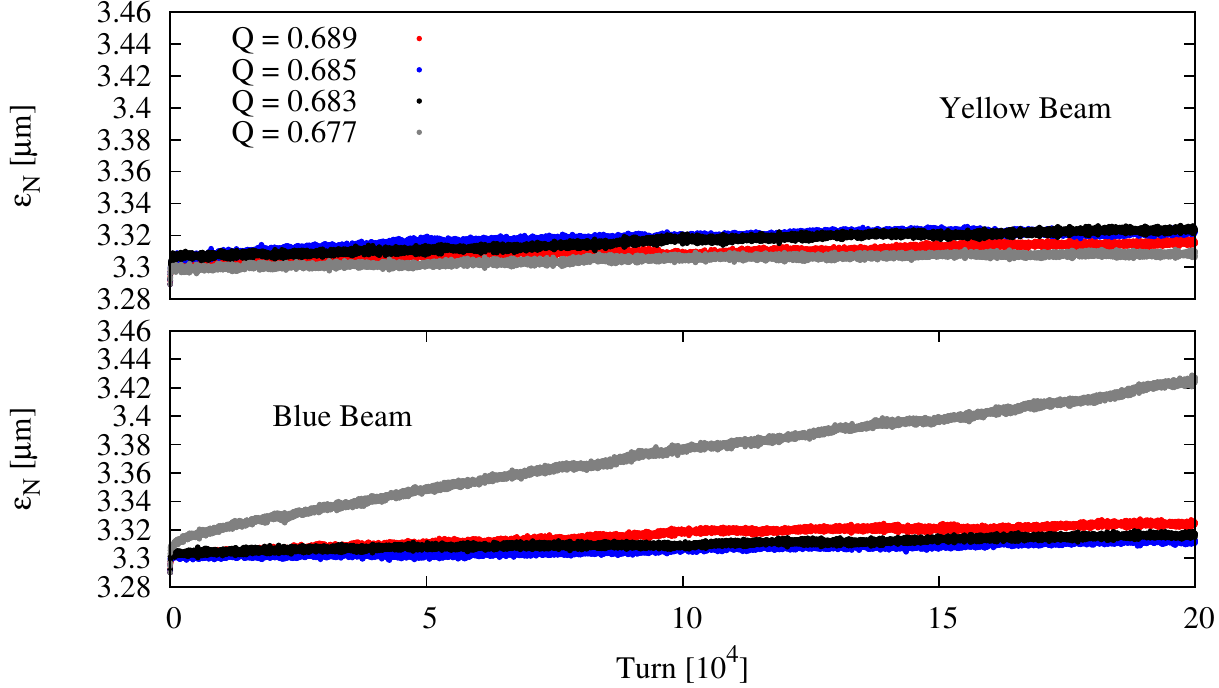}
\end{center}
\caption{Simulated emittance during the tune scan.}
\label{emit_qscan}
\end{figure}

The tune scan was reproduced in numerical simulations. Figure \ref{fft_qscan} shows the FFT of the centre of motion during the the tune scan. It is observed
that even when the $\pi$-mode is on top of the 2/3 resonance it remains stable. Figure \ref{emit_qscan} shows the vertical emittance growth for both beams.
Only the Blue beam, pushed towards the resonance, is affected and a blow-up is observed only when the beam--beam tune spread overlaps the resonance. This
is very consistent with experimental data and confirms that the 2/3 resonance is not a concern for beam stability. The only non-linear element in the model
is the beam--beam interaction. The absence of sextupoles explains why the stop band of the resonance is narrower in the numerical simulation.

\section{Coherent mode suppression}

Coherent beam--beam mode suppression has been investigated in Refs. \cite{split,resonance}, where it was shown that the following techniques can be used to damp
the modes:

\begin{itemize}
 \item Phase advance adjustment between colliding IPs
 \item Synchro--betatron coupling. If $\xi\approx Q_s$, the $\pi$-mode can be damped by the sidebands of the continuum
 \item Beams colliding on different working points (tune split)
\end{itemize}

Although these effects could all be reproduced in simulations, one has to consider the constraints associated with the machine layout and beam
parameters. Due to the magnet powering scheme in the RHIC there is very little flexibility to adjust the phase advance between the colliding IPs (IP6 and IP8). The
synchrotron tune $Q_s$ is of the order of $5.0\times10^{-4}$, which is much smaller than the expected beam--beam parameter
in the presence of head-on compensation ($\xi\approx 0.02$--0.03) making it impossible to profit from synchro--betatron
coupling. This leaves the tune split as the only option for coherent mode suppression in the RHIC.

\begin{figure}[htb]
\begin{center}
\includegraphics[width=0.45\textwidth]{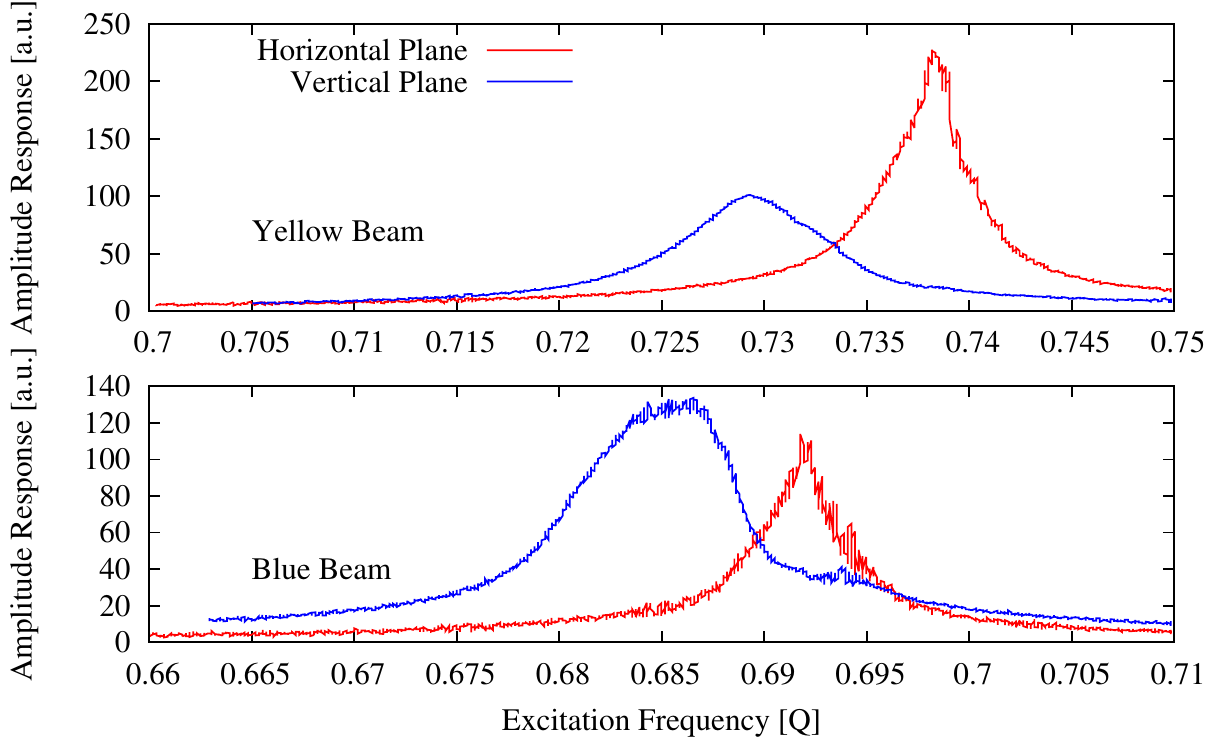}
\end{center}
\caption{BTF measurement with split tunes.}
\label{split_BTF}
\end{figure}

To fully suppress the coherent modes the tune split between the two beams has to be larger than the beam--beam parameter, in which case the coherent modes
will cluster inside the incoherent continuum and experience Landau damping \cite{split}. This can be achieved at the RHIC with tunes of about (0.695, 0.685) for the
Blue beam and (0.74, 0.73) for the Yellow beam. Figure \ref{split_BTF} shows an example of a BTF measurement with the beams on different tunes. The coherent
modes are completely suppressed, as expected from the theory.

\begin{figure}[htb]
\begin{center}
\includegraphics[width=0.45\textwidth]{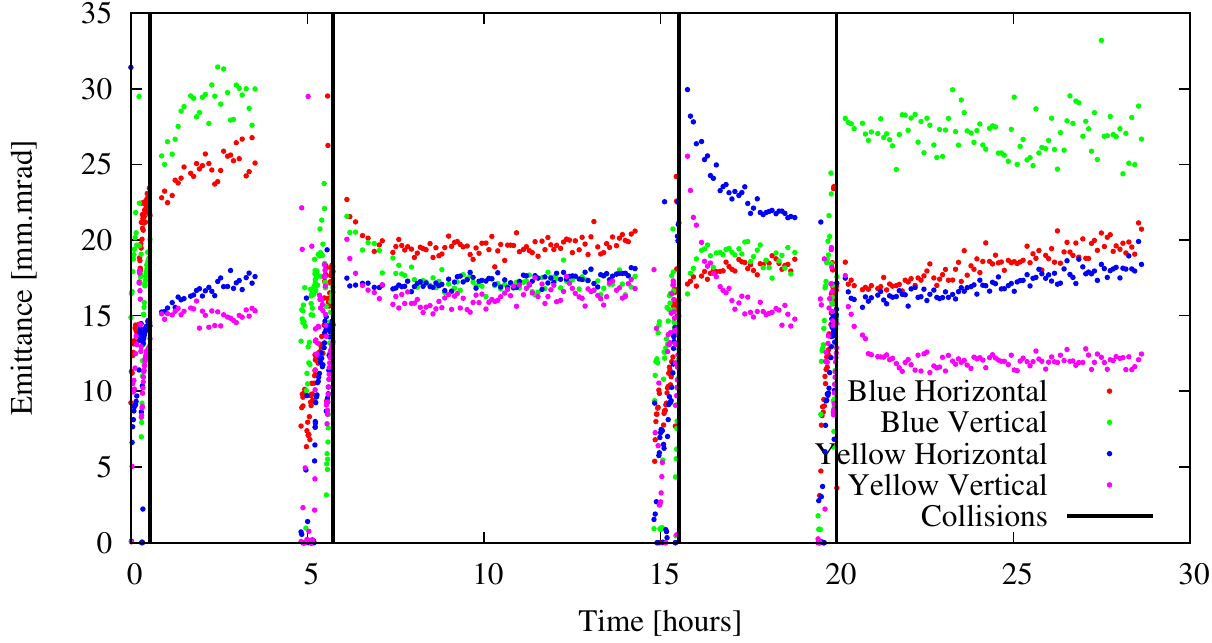}
\end{center}
\caption{Emittance measurements during fills with split tunes.}
\label{split_emit}
\end{figure}

Figure \ref{split_emit} shows emittance measurements over four consecutive stores with split tunes. A strong emittance blow-up is observed in three
fills out of four as soon as the beams are brought into collision, leading to poor luminosity performance. For comparison, the emittance at the beginning of the stores is generally
around 15\,mm\,mrad for normal operation. This behaviour for colliding beams with unequal tunes had been predicted in past simulations and theoretical analysis \cite{resonance},
where it was stated that operating a collider with unequal tunes could lead to coherent beam--beam resonance excitation and, providing the modes lie inside the incoherent continuum,
emittance blow-up.

\begin{figure}[htb]
\begin{center}
\includegraphics[width=0.45\textwidth]{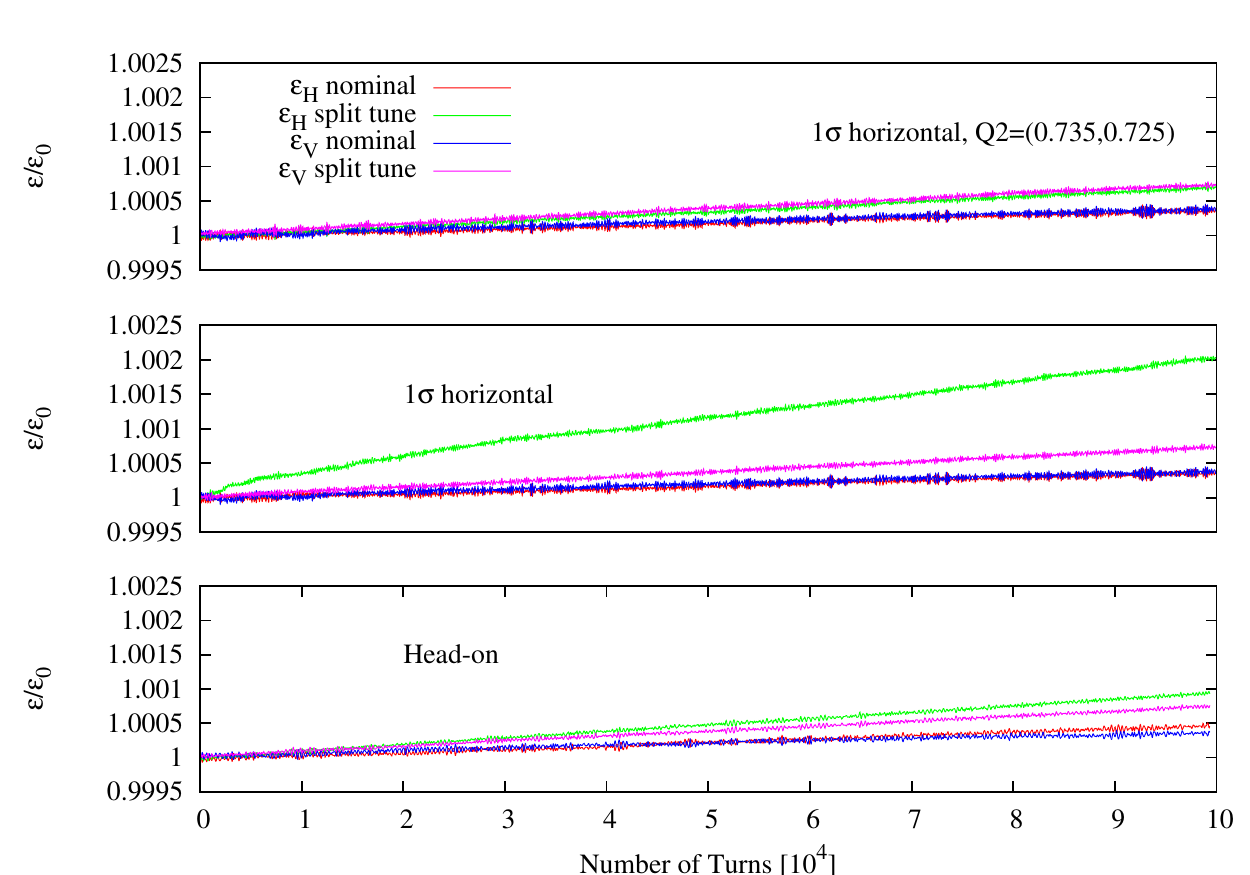}
\end{center}
\caption{Simulated emittance growth in the vicinity of the resonance for head-on interactions (bottom), separated beams (middle) and separated beams away from the resonance (top).}
\label{split_emit_sim}
\end{figure}

Using the estimated working points (0.689, 0.691) and (0.74, 0.73) and beam--beam parameter (0.013) one can compute the frequencies of the coherent modes using a rigid bunch model.
In this specific case the machine was operated in the vicinity of a resonance of the form $4Q_1-Q_2$, which is excited by offset collisions. Numerical simulations were
carried out to assess the impact of this resonance on emittance. The results of these simulations are shown in Fig. \ref{split_emit_sim}, where three cases were considered:

\begin{itemize}
 \item Head-on collisions with tunes close to the resonance (experimental conditions)
 \item Collisions with an offset of 1$\sigma$ in the horizontal plane with tunes close to the resonance
 \item Collisions with an offset of 1$\sigma$ in the horizontal plane with tunes away from the resonance
\end{itemize}

A strong emittance blow-up is observed in the case of offset collisions with working points close to the resonance condition. When the beams are colliding head-on
or the working points are moved away from the resonance the conditions simulated for equal tunes are almost recovered. Simulations appear to confirm the hypothesis of
a coherent beam--beam resonance of odd order. We could expect that by properly setting the working points to avoid resonances nominal luminosity performance could be
achieved. Another important parameter in the RHIC is the polarization. During the split tune experiment a very poor polarization was measured for the Yellow beam (0.74, 0.73),
ruling out the possibility of running the RHIC in this configuration.

\begin{figure}[htb]
\begin{center}
\includegraphics[width=0.45\textwidth]{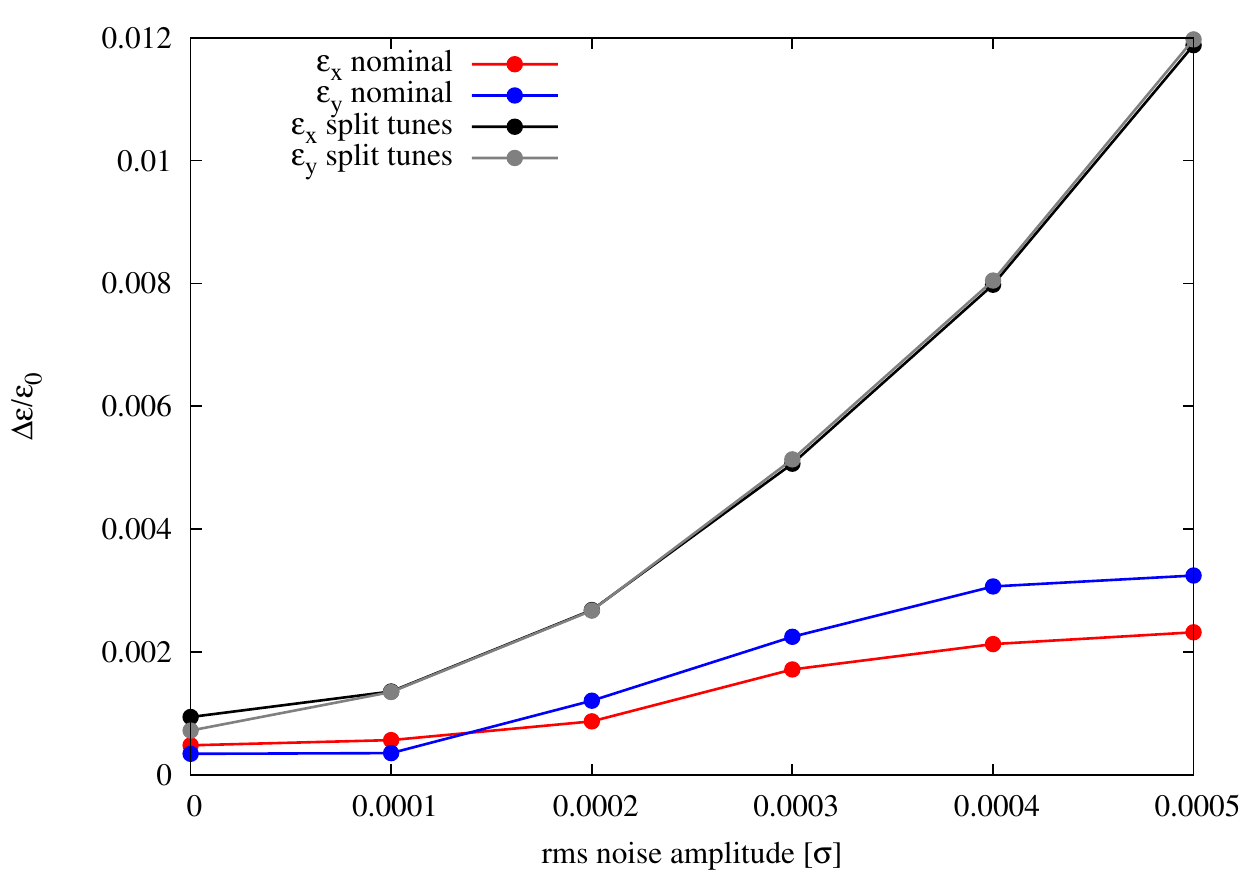}
\end{center}
\caption{Emittance growth in the presence of white noise for colliding beams with equal and unequal tunes.}
\label{emit_wn}
\end{figure}

Figure \ref{emit_wn} shows the emittance growth due to white-noise excitation. In this case, beam parameters were set to be away from any low-order resonance.
It is clearly observed that colliding the beams with unequal tunes degrades the situation and makes the beams more sensitive to external excitation. This was
not verified experimentally and would need confirmation but could become an issue if operation with split tunes is considered for a collider.

\section{Machine impedance}

Head-on compensation with electron lenses will significantly reduce the beam--beam tune spread and Landau damping. The interplay with machine impedance
was studied in numerical simulations using the RHIC impedance model which takes into account the contribution of stripline BPMs (Beam Position Monitor), bellows and resistive wall.

\begin{figure}[htb]
\begin{center}
\includegraphics[width=0.45\textwidth]{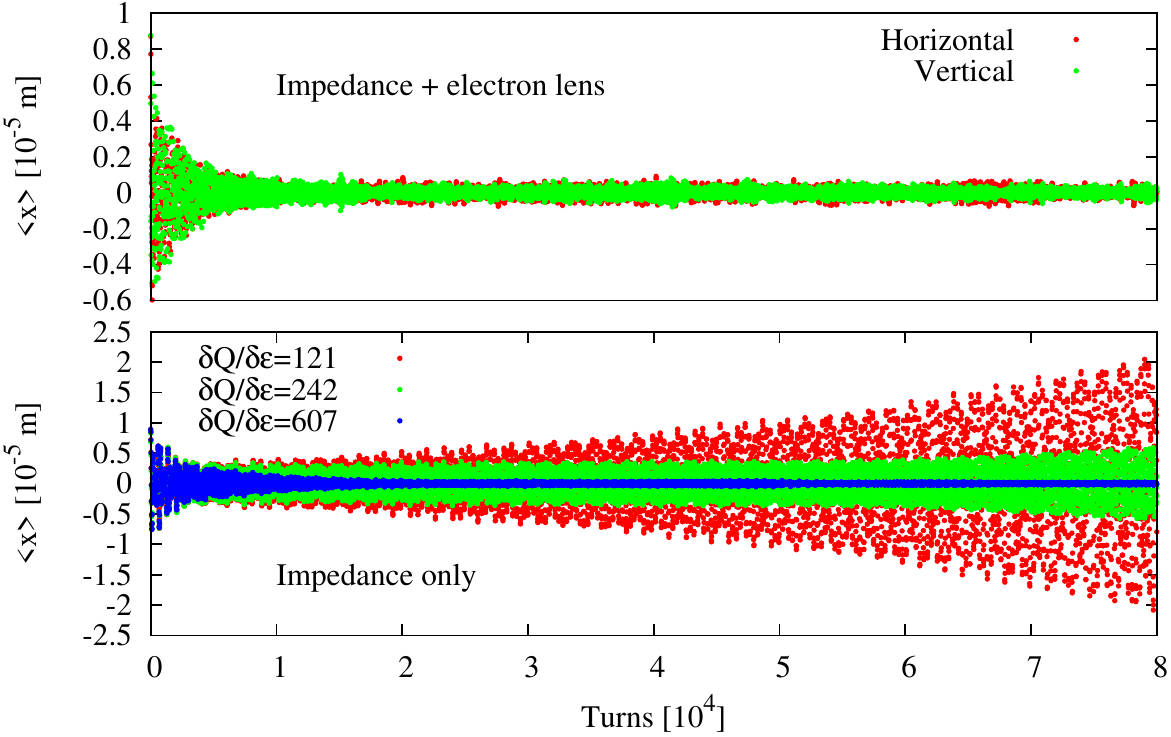}
\end{center}
\caption{Beam stability with machine impedance only and including electron lens for $Q'=2.0$.}
\label{imp}
\end{figure}

\begin{table}

\caption{Stabilizing detuning coefficients derived from tracking simulation with impedance and RHIC non-linear model.\label{tab:dq}}
\begin{tabular}{cccc}
\hline\hline
                           &$\partial Q_x / \partial \epsilon_x$&$\partial Q_y / \partial \epsilon_y$&$\partial Q_y / \partial \epsilon_x$\\
                           & m$^{-1}$ & m$^{-1}$ & m$^{-1}$ \\ \hline
  Tracking                 &  607 & 607 &  417             \\
  Non-linear model         &  314 & 387 &  463            \\
\hline\hline
\end{tabular}
\end{table}

Figure \ref{imp} shows the results of simulations for a chromaticity of 2.0, which corresponds to what is generally used in regular RHIC operation. The bottom plot
is a scan in octupolar detuning with impedance only. Stability is achieved for a detuning coefficient $\delta Q_x/\delta\epsilon_x=607$\,m$^{-1}$. Table \ref{tab:dq} summarizes
the stabilizing coefficient obtained from tracking simulations and the detuning coefficients derived from the RHIC non-linear model.
Considering the uncertainties from the impedance model and the difficulty in accurately computing the stability threshold from tracking simulation, it is not unlikely that the
machine non-linearities provide sufficient detuning to stabilize the beam. This would be consistent with the fact that instabilities are generally not observed during
RHIC polarized proton runs. Even when half-compensated the beam--beam non-linearities will provide significantly larger detuning than the results from Table \ref{tab:dq}.
Machine impedance is therefore not considered to become an issue for stability. This was confirmed by simulations as shown in the top plot of Fig. \ref{imp}, where the beam
is fully stable with electron lenses running at half compensation.

\section{Electron lens driven TMCI}

When a proton bunch interacts with the electron beam it will drive Larmor oscillations of the electrons along the interaction region resulting in an $s$-dependent kick onto
the proton bunch. This can be interpreted as an electron lens impedance comparable to or larger than the machine impedance. Its strength depends on the electron lens parameters and under
certain conditions can lead to transverse mode coupling instabilities (TMCI). This effect was studied in detail in Ref. \cite{TMCI_elens}, where it was shown that the $s$-dependent
momentum change of the protons can be modelled with the following wake function:

\begin{equation}
 \Delta p_{x} = W[\Delta x \sin(ks)+ \Delta y (1-\cos(ks))],
\end{equation}
where $\Delta x$ and $\Delta y$ are the offsets of the source in the horizontal and vertical planes respectively and $W$ is a constant depending on both the beam--beam
parameters of the electron and proton beams and the solenoid field $B$. A similar equation is also valid for the vertical plane $y$. The variable $k$ is defined as

\begin{equation}
 k = \frac{\omega_L}{(1+\beta_e)c},
\end{equation}
where $\beta_e$ is the relativistic $\beta$ of the electron beam, $c$ is the speed of light and $\omega_L$ is the Larmor angular frequency defined as

\begin{equation}
 \omega_L=\frac{eB}{\gamma_e m}.
\end{equation}

Using this wake function and considering uniform and equal transverse distributions for the proton and electron beams, it is possible to analytically derive the
TMCI threshold and hence the required solenoid field to ensure stability. This threshold can be expressed expressed as \cite{TMCI_elens}

\begin{equation}
B_{\mathrm{th}} = 1.3 \frac{e N_p \xi_e}{r^2 \sqrt{\Delta Q Q_s}},
\label{bth}
\end{equation}
where $N_p$ is the proton bunch intensity, $\xi_e$ is the electron lens beam--beam parameter, $r$ is the radius of the beam ($r\approx2\sigma$ for a Gaussian distribution),
$\Delta Q$ is the separation between horizontal and vertical tunes and $Q_s$ is the synchrotron tune. Using typical RHIC parameters ($N_p = 3.0\times10^{11}$ protons per bunch,
$\xi_e=0.011$, $\Delta Q = 0.01$, $Q_s=5.0\times10^{-4}$ and $r\approx0.8$\,mm), a threshold field of 14\,T is found, which is approximately a factor 2 above the design field of
6\,T. This six-dimensional electron lens interaction was built into the code BeamBeam3D to study beam stability with an electron lens in multi-particle tracking simulations.
Benchmarking with theoretical predictions was done using linearized beam--beam kicks, which allows for direct comparison.

\begin{figure}[htb]
\begin{center}
\includegraphics[width=0.45\textwidth]{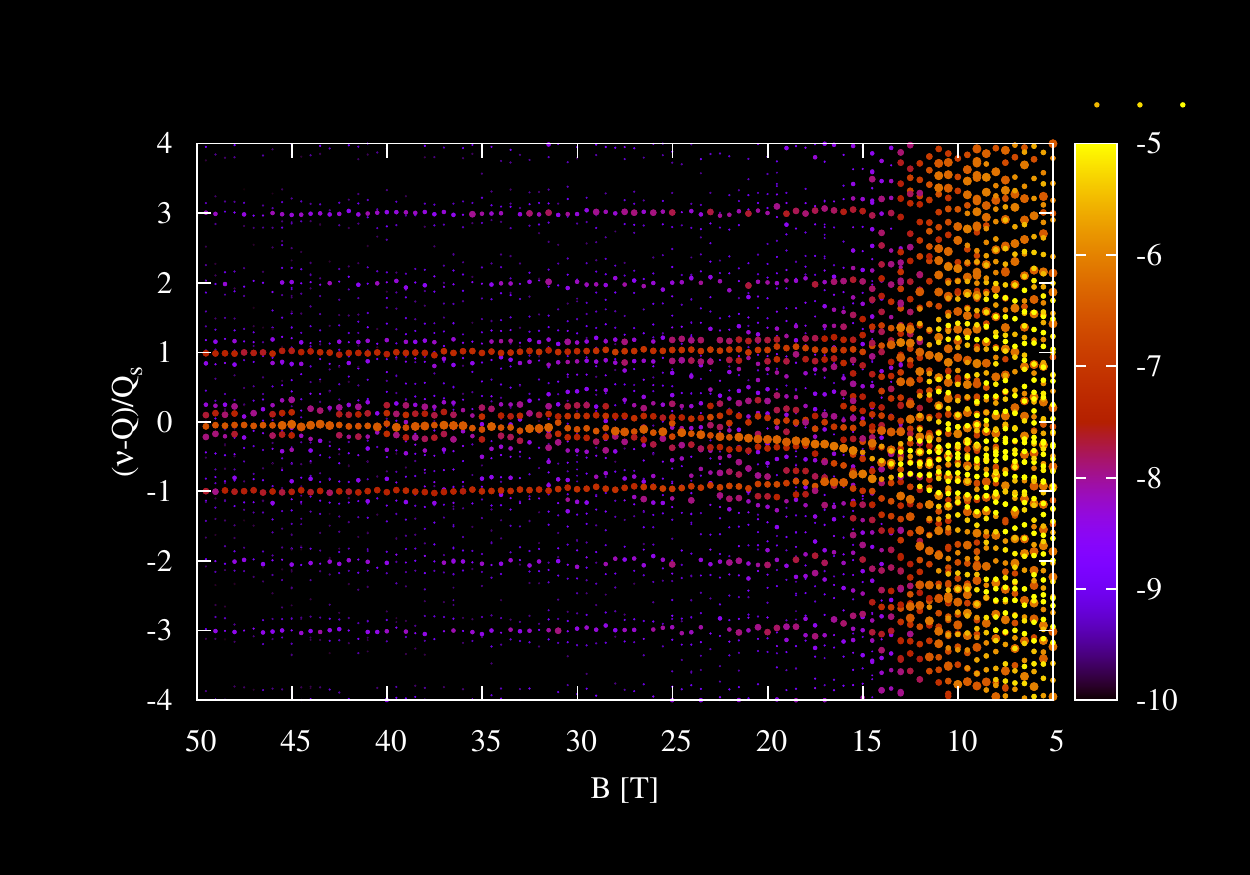}
\end{center}
\caption{Synchro--betatron mode frequencies and amplitudes (colours) as a function of the solenoid field and using a linearized model
(no Landau damping). The transverse mode coupling instability occurs at around 14\,T.}
\label{el_only}
\end{figure}

Figure \ref{el_only} shows the results of this benchmarking using the same beam parameters as the field threshold computation from Eq.\,(\ref{bth}).
The transverse mode coupling instability occurs at around 14\,T, which is consistent with theoretical expectations. These results only include interactions with an electron
lens; in the presence of beam--beam (proton--proton) interactions coherent motion is driven by these additional interactions and the mode frequencies are modified. This
is especially true in the presence of strong synchro--betatron coupling from the beam--beam interaction, which, in the case of the RHIC, is a result of the hourglass effect
($\beta^*/\sigma_s\approx1$, no crossing angle).

\begin{figure}[htb]
\begin{center}
\includegraphics[width=0.40\textwidth]{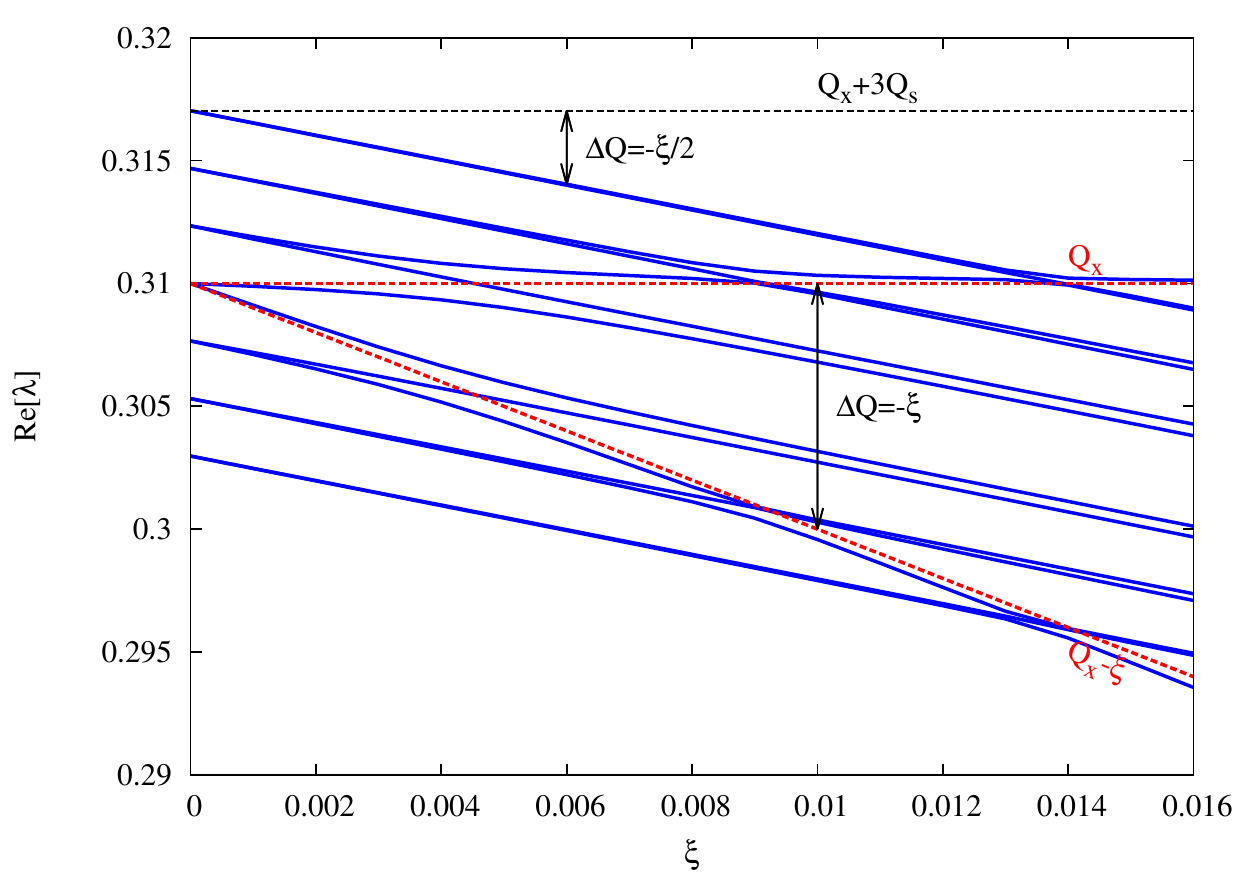}
\end{center}
\caption{Synchro--betatron mode frequencies as a function of the beam--beam parameter for $Q_s\approx0.0025$ and $\beta^*/\sigma_s\approx1$.}
\label{bb_sbm}
\end{figure}

This is illustrated in Fig.\,\ref{bb_sbm}, where it is clearly seen that the synchrotron sidebands are deflected by the beam--beam $\pi$ and $\sigma$ modes. In this
case the mode frequencies were computed using a linearized model based on the circulant matrix approach \cite{circ_mat}, in which case the Yokoya factor is equal to 1.

\begin{figure}[htb]
\begin{center}
\includegraphics[width=0.45\textwidth]{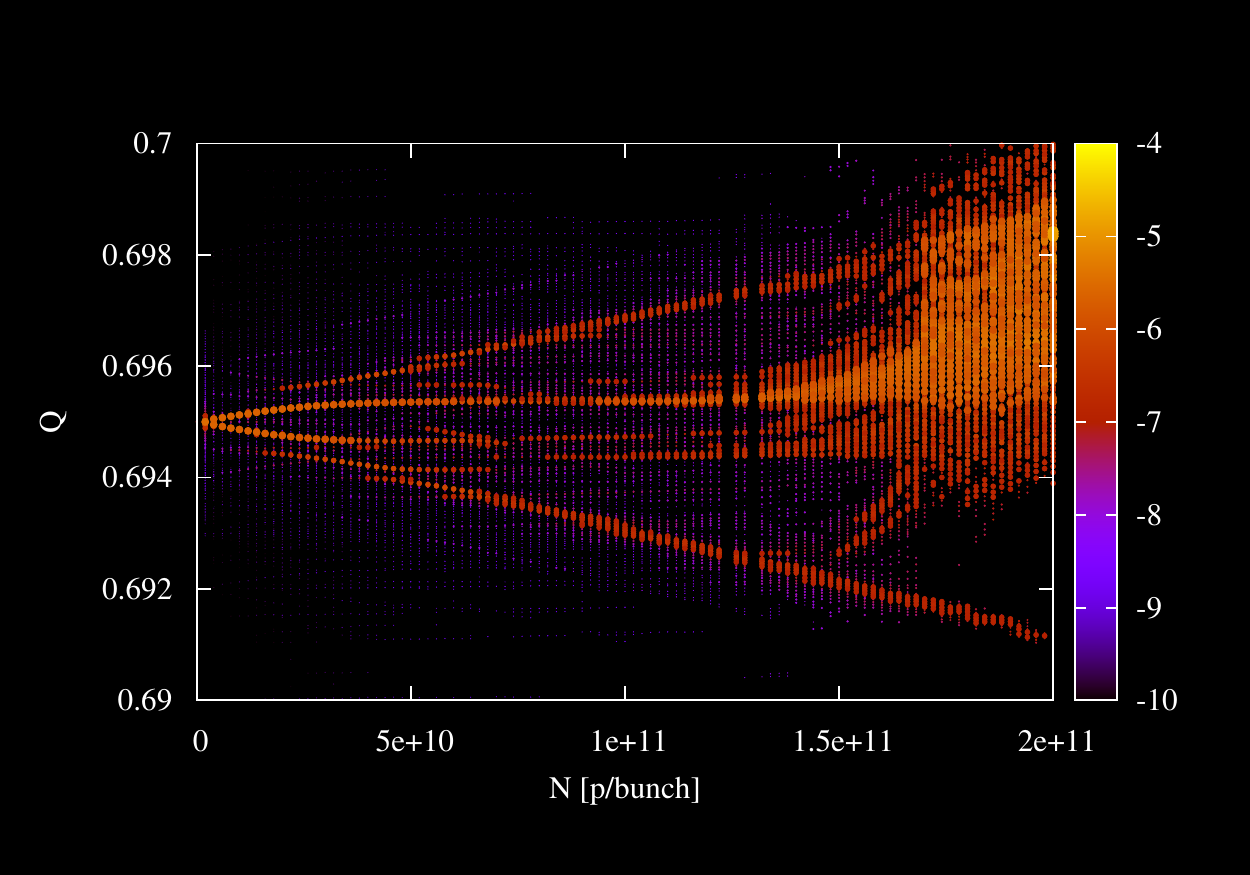}
\end{center}
\caption{Synchro--betatron mode frequencies and amplitudes (colours) as a function of bunch intensity for a solenoid field of 6\,T and  using a linearized model
(no Landau damping). The transverse mode coupling instability occurs at around $1.0\times10^{11}$ protons per bunch.}
\label{bb_el_sbm}
\end{figure}

Figure \ref{bb_el_sbm} shows tracking results including beam--beam and electron lens using linearized beam--beam kicks and RHIC beam parameters. The expected threshold
from Eq.\,(\ref{bth}) is $2.0\times10^{11}$ in this case. The mode coupling instability in the presence of coherent beam--beam effects is reduced by a factor 2 for these
parameters with respect to theoretical expectations without coherent beam--beam effects (the threshold in terms of solenoid field scales with $N_p^2$).

The above simulations were carried out using a linearized model which does not include the amplitude detuning related to the non-linearities of the beam--beam force and
hence its contribution to Landau damping. Landau damping could provide additional stability and mitigate the electron lens driven TMCI. In order to include this effect,
we carried out tracking simulations using the full non-linear beam--beam force. The proton--proton interactions are computed using a Poisson solver, making no assumption on the
beam distribution, while the interaction with the electron lens is done assuming elliptical Gaussian shapes but allowing for a tilt angle of the phase-space distribution in
order to account for the coupling introduced by the solenoid field; more detailed studies would be required to verify the validity of the Gaussian approximation and its impact
on Landau damping. The proton bunch is sliced longitudinally into 50 slices, which correspond to 10 times the wavelength of the Larmor oscillations. Although the high-frequency
component of the wake function should not have a significant impact on stability, it is necessary to perform systematic studies regarding the effect of the number of slices. This may
introduce some aliasing issues when sampling the electron oscillations and eventual smoothing approximations could apply. The importance of these parameters is under investigation and
will not be covered in detail in this paper.

\begin{figure}[htb]
\begin{center}
\includegraphics[width=0.45\textwidth]{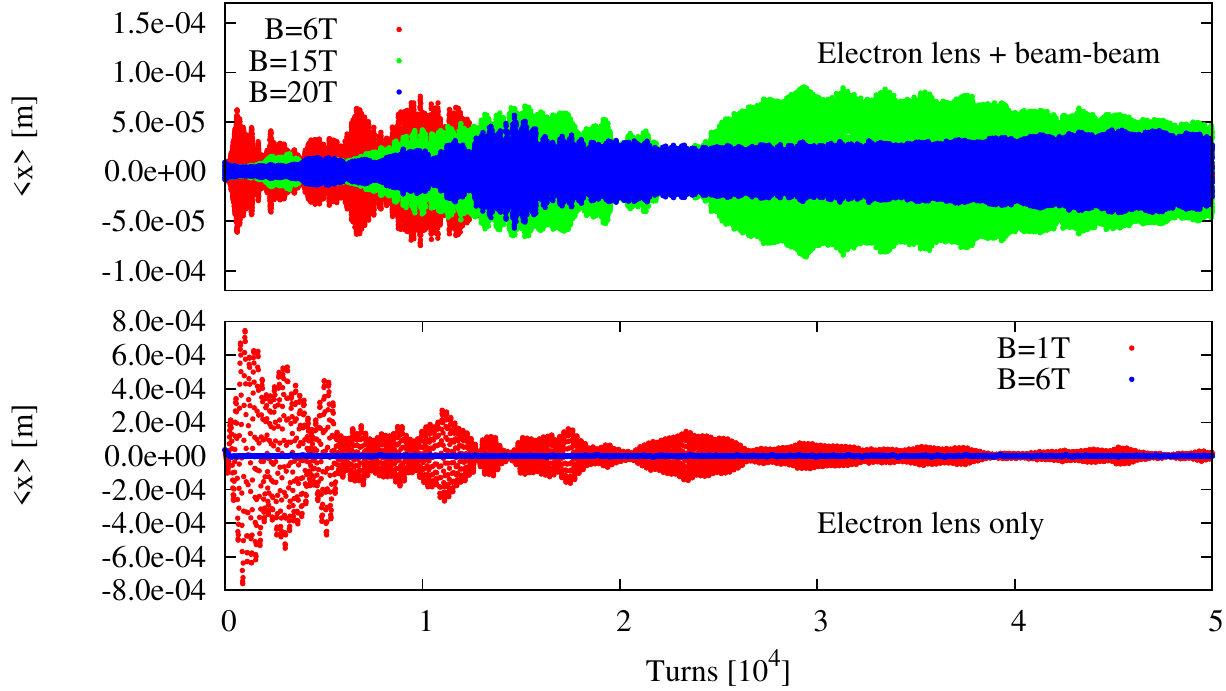}
\end{center}
\caption{Solenoid field scan including non-linear beam--beam force with (top) and without (bottom) proton--proton interactions.}
\label{bb_el_ld}
\end{figure}

Figure \ref{bb_el_ld} shows the results of a solenoid field scan including Landau damping for cases with and without proton--proton interactions. The case without proton--proton
interactions at the bottom provides a direct comparison with theoretical predictions and illustrates the impact of Landau damping. The theoretical threshold was estimated to be
approximately 14\,T. Including Landau damping, this threshold is significantly reduced and stability is achieved for a solenoid field between 1\,T and 6\,T, which is within
the RHIC electron lens design. Unfortunately, the degradation due to coherent beam--beam effects is also observed in the presence of Landau damping and the beam could not be
stabilized for a field up to 20\,T, which is well above design.

\begin{figure}[htb]
\begin{center}
\includegraphics[width=0.45\textwidth]{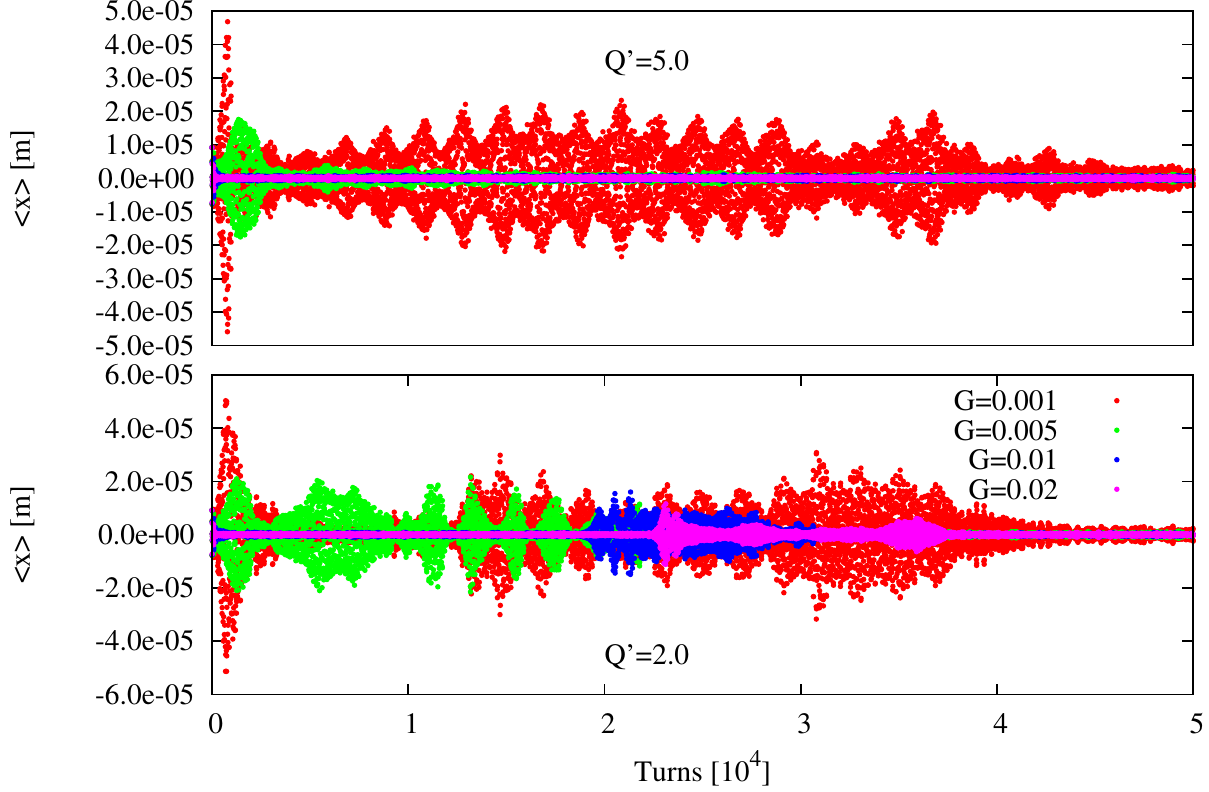}
\end{center}
\caption{Rigid bunch damper gain scan with $Q'=2.0$ (bottom) and $Q'=5.0$ (top).}
\label{bb_el_damper}
\end{figure}

As demonstrated in Ref. \cite{damper}, the combination of a strong rigid bunch damper and relatively high chromaticity can provide stability to higher order head--tail modes ($m>0$)
and could even push the TMCI to higher threshold. Figure \ref{bb_el_damper} shows the results of tracking simulations including a rigid bunch damper and chromaticity. At $Q'=2.0$,
which corresponds to normal RHIC running chromaticity, a clear improvement is observed when the damper gain is increased but the beam remains unstable even at high gain. Increasing
the chromaticity makes the damper more efficient against the mode coupling instability and the beam can be stabilized with a damper gain of 100 turns.

\section{Conclusions}

Head-on beam--beam compensation with electron lenses reduces the beam--beam tune spread allowing one to accommodate larger bunch intensity at the current RHIC working point.
In return, the contribution from the beam--beam tune spread to Landau damping is reduced while the coherent beam--beam modes remain unaffected.
This may have some detrimental effects on beam stability and luminosity performance. Beam experiments were conducted at the RHIC to understand the impact of coherent beam--beam effects on beam dynamics:

\begin{itemize}
 \item \textbf{Impact of the 2/3 resonance}: the beam--beam $\pi$-mode was driven onto the 2/3 resonance without effect on beam stability or cross talk between beams. Losses and emittance
blow-up were observed in the beam moving towards the resonance, which is attributed to incoherent effects when the tune spread overlaps the resonance stop band.
This is consistent with theoretical estimates and tracking simulations
 \item \textbf{Coherent mode suppression}: coherent beam--beam mode suppression with tune split was attempted. This resulted in significant luminosity performance degradation
due to emittance blow-up when bringing the beams into collision. This effect could be attributed to the excitation of coherent beam--beam resonance as predicted in Ref. \cite{resonance}. Tracking
simulations also support this hypothesis
\end{itemize}

Numerical simulations were carried out to understand possible limitations coming from machine impedance and electron lens impedance. It was shown that the intrinsic machine non-linearities
provide almost sufficient detuning to stabilize instabilities driven by machine impedance. This is consistent with experimental data, as instabilities are generally not observed in regular
operation. The electron lenses are foreseen to compensate for only half of the full beam--beam tune spread. The remaining tune spread would still be significantly larger than the simulated
stabilizing octupolar detuning, leading to the conclusion that machine impedance is not a limitation for operation with electron lenses.
Another aspect investigated in this paper is the electron lens driven TMCI. It was found that the RHIC design field is not sufficient to ensure stability with the current machine layout
and beam parameters. A possible solution to overcome this issue would be the implementation of a bunch-by-bunch transverse damper combined with slightly higher than nominal chromaticities.
These results are preliminary and more systematic studies and model refinements are required to draw final conclusions. The electron lens driven TMCI could also be mitigated using beam
parameters such as the distance between the horizontal and vertical tunes or the $\beta$-function at the electron lens. These alternative solutions should be investigated in future studies.

\section{Acknowledgments}

The authors would like to thank J. Qiang for his help regarding the implementation of the RHIC lattice and
electron lenses into the code BeamBeam3D, the RHIC operation and M. Bai and A. Marusic for their support
with data acquisition.

\end{document}